\documentstyle[12pt]{article}
\begin{document}
\pagestyle{empty}
\rightline{LPT Strasbourg 95-20}\par
\begin{center}
 {\large \bf Local Fractional Supersymmetry for Alternative Statistics }
\end{center}

 \vskip 1 truecm

\begin{center}
    N. Fleury$^*$  and M. Rausch de Traubenberg  \\

 Laboratoire de Physique Th\'eorique, Universit\'e Louis Pasteur \\
  3-5 rue de l'universit\'e, 67084 Strasbourg Cedex, France \\
  and \\
  Centre de Recherches Nucl\'eaires, B\^at 40/II 67037 Strasbourg Cedex 2 \\
\end{center}

\vskip 1 truecm

\begin{center}
  $^*$ IUFM d'Alsace- 200 route de Colmar, 67000 Strasbourg
\end{center}

\vskip 1 truecm
\abstract{A group theory justification of one dimensional
fractional supersymmetry
is proposed using an analogue
  of a coset space, just like
 the one introduced in $1D$ supersymmetry. This theory is then gauged to
obtain a local fractional supersymmetry {\it i.e.} a fractional
supergravity which is then quantized {\it \`a la Dirac} to obtain
an equation of motion for a particle which is in a representation of
the braid group and should describe alternative statistics.
A formulation invariant under general reparametrization is   given,
by means of a curved fractional superline.}

\eject
 \pagestyle{plain}
\pagenumbering{arabic}
\vskip .5 truecm

With string theory\cite{gsw}, a new approach came out in the description of
space-time symmetry. Indeed, by studying the symmetries on the world sheet
of the string one can get the space-time properties of the
string states, {\it i.e.}
the particles (representations of the gauge group  are controlled by Kac-Moody
algebra\cite{go} and of the Poincar\'e one by (su\-per)conformal
invarian\-ce\cite{bpz}).
However,
all those results were anticipated and an alternative formulation of
relativistic wave equations\cite{wl1,wl2,wl3}
 and  quantum field theory can be obtained
with the study of physics on the world line of the particle. Particles
with spin $N/2$ could be described by an  $N-$ extented supersymmetry\cite{wl3}
 on the
world line, and gauge symmetries by the introduction of internal Grassmann
variables\cite{wlg}. All this was recently promoted into  an alternative  and
 efficient description of field theory using the world-line
 formalism\cite{wlf}, introducing
$1D$ Feyman rules and appropriate one dimensional Green
func\-tions\cite{wlfey}.

However, the spin statistics theorem and the Haag, Lopuszanski and Sohnius
no-go theorem\cite{no-go} tell us that supersymmetry is the more general
 non-trivial
symmetry that one can  consider; as soon as we are in a $D \le 3$ dimensional
space-time one can find statistics that are neither fermions nor bosons,
but anyons\cite{anyons} or particles which admit fractional statistics.
Technically
the former particles are in the representation of the permutation group
and the latter of the braid group. In the meantime some extensions
of $1D$ supersymmetry have been considered,  for instance
parasupersymmetry\cite{psusy1,psusy2} or fractional
supersymmetry\cite{fsusy1,fsusy2,fsusy3,fsusygr,fsusy4}. It has been proved
that
 $1D$ parasupersymmetry of order $p$ could be equivalent to  $p-$extended
world-line supersymmetry and describes particles of spin
${p \over 2}$\cite{psusy2}.

Fractional supersymmetry has been recently the subject of intensive
studies\cite{fsusy1,fsusy2,fsusy3,fsusygr,fsusy4}.
Following the way which leads from $1D$ supersymmetry to the Dirac equation,
applied in the context of fractional supersymmetry, we get a new equation
 acting on states which are in the representation of the braid group.
This equation can be seen as an extension of the Dirac equation in the sense
that the $n-$th power of  the field operator  is equal to the Klein-Gordon
one. In this paper we particularize the case $n=3$. In a first step
we define, in analogy with  the superspace, the fractional superspace as
some kind of coset space reobtaining all what has been done in the framework
of fractional susy. In a second step, we construct a local fractional
supersymmetry {\it i.e} a fractional supergravity by using
two one dimensional gauge fields: the einbein and a field which can be compared
to the $1D$ gravitino and that we call the fractional gravitino.
A formulation, in a curved fractional superline, which is invariant under
general coordinate transformations  is then given.
The second part is devoted to the quantization of the theory, taking under
consideration  the first and second class constraints\cite{Dirac}. After having
constructed the Fock space with the help of the $q-$ deformed
oscillators\cite{q-osc.}
we obtain a new equation, that  we call the fractional Dirac
equation.\\

 \noindent
 {\bf  I.Fractional Superspace and Fractional Supersymmetry}\\

Historically, 4D supersymmetry has been built explicitly, components by
components
(see for example \cite{susy}). However it was understood later  that this
symmetry is just a consequence of a  symmetry in a so-called superspace which
can be seen as the coset space of the Superpoincar\'e group by the Lorentz
group\cite{susy}.  The  superspace is just the 8-fold space $(x^\mu,\theta)$,
where $x^\mu$ is the space-time components and $\theta$  its spinor partner.
 Because
we are studying physics on the world line   we just particularize the
$1D$ case. Noting $H$ the generator of the time translation, and $Q$ the
generator of the susy transformation, a point $(t,\theta)$ is parametrized by

\begin{equation}
exp(t H + \theta Q).
\end{equation}

\noindent
Using the susy algebra $ [H,Q] = 0, \{Q,Q\} = -2 H $ and the definition of
a susy transformation with parameter $\epsilon$ we get the transformation law

\begin{eqnarray}
exp(t' H + \theta' Q) &=& exp( \epsilon Q) exp(t H + \theta Q)
                        \nonumber \\
                  &=&   exp((t+i\epsilon \theta) H +(\theta + \epsilon) Q).
\end{eqnarray}

After having  introduced the superfield
$ \Phi(t,\theta) = x(t) + i\theta \psi(t)$, it becomes easy to construct the
susy transformation  on the fields themselves and to build an invariant
action.\\

Supersymmetry is the only nontrivial $Z_2$-extension of the Poincar\'e
algebra\cite{no-go}
which is not in contradiction with the spin statistics  theorem. However, in
one dimension there is no obstruction to build other non trivial extensions.
This is for instance parasupersymmetry\cite{psusy1,psusy2}
or fractional supersymmetry\cite{fsusy1,fsusy2,fsusy3,fsusygr,fsusy4} .
The latter possesses a  $Z_n$-structure, and
 through
this article we will concentrate only on the $n=3$ case. Group theory
justification of fractional supersymmetry (fsusy) has been undertaken
in\cite{fsusygr}
but without the introduction of the analogue of the superspace, we call the
fractional superspace (fsuperspace). A point in a fsuperspace is given by
$(t,\theta)$, where $\theta$ is a {\it real} generalized Grassmann
variable\cite{gca1,gca2,gca3} of grade one\footnote{
Some confusion exists between generalized Grassmann variables and
Paragrassmann ones. Although, in the case of one variable those
algebras coincide,
they are different in general. The latter appears in the frame of
parastastistics
and is in some representation of the permutation group \cite{para}, whereas
the former
is just in a representation of the braid group\cite{gca1,gca2,gca3} .}
submitted to the constraint $\theta^3 = 0$.

\noindent
Let  $Q$ be the generator of fsusy satisfying the condition

\begin{equation}
Q^3 = - H,
\end{equation}

\noindent
and define a point in the fsuperspace by its parametrization

\begin{eqnarray}
exp_{gr}(t H + \theta Q)  &=& exp_{gr}(t H) exp_{gr}(\theta Q )\nonumber \\
                        &=& exp(t H) exp_{q} ( \theta Q ),
\end{eqnarray}

\noindent
where $exp_{gr}$ is the graded exponential ($t$ is of grade zero and
$\theta$
 of grade one), $q$ is a pri\-mitive cub\-ic root of unity that we can
take e\-qual to
 $exp(2i\pi/3)$ without losing   generality and

\begin{equation}
exp_{q^a} (x) = \sum\limits_{k=0}^\infty {x^n \over \{n\}_a!},
\end{equation}
where  $ \{n\}_a!=\{n\}_a \{n-1\}_a \dots \{1\}_a,
 \{k\}_a= {1-q^{ak} \over 1-q^a}$.\\

\noindent
This series exactly stops with its $(n-1)-$th power because $\theta^n = 0$,
in the general case. For
$n=2$ we have  only two terms and in this case the usual exponential
coincides exactly with the q(=-1)-exponential and we recapture the
 definition (2).
Going back to $n=3$ we get

\begin{equation}
exp_{q} (\theta Q) = 1 + \theta Q -q (\theta Q )^2.
\end{equation}

\noindent
Now introduce $\epsilon$ the {\it real} parameter of the fsusy transformation
($\epsilon^3=0$), and
using the q-mutation relations (see the appendix for the justification of
the q-mutators)

\begin{eqnarray}
Q \theta &=& q^2 \theta Q \nonumber \\
Q \epsilon &=&q^2 \epsilon Q \\
\theta \epsilon &=& q \epsilon \theta, \nonumber
\end{eqnarray}

\noindent
we get the fsusy transformation in the fsuperspace

\begin{eqnarray}
exp(t' H) exp_{q} (\theta' Q) &=&
                    exp_{q}( \epsilon Q) exp(t H) exp_{q}( \theta Q) \\
 &=& exp((t+q(\epsilon^2 \theta + \epsilon \theta^2))H)
  exp_{q}((\theta + \epsilon) Q). \nonumber
\end{eqnarray}

\noindent
The transformations we have obtained coincide exactly with those
of\cite{fsusygr}.
It has to be stressed again that $t'=t+q(\epsilon^2 \theta + \epsilon
\theta^2)$
is real, as it should be. To obtain this equation we just developed explicitly
eq.(8).  The next step, to build an action, is to
introduce a {\it real} fractional superfield $\Phi$ (fsuperfield)
belonging to the
fsuperspace.  The   Taylor expansion of  $\Phi(t,\theta)$  gives

\begin{equation}
\Phi(t,\theta) = x(t) + q^2 \theta \psi_2(t) + q^2 \theta^2 \psi_1(t),
\end{equation}

\noindent
where $x(t), \psi_1(t), \psi_2(t)$ are three {\it real} fields respectively
of grade $0,1,2$ such that $\psi_1^3=\psi_2^3=0$ and are
submitted to the q-mutation relations (see the appendix)

\begin{eqnarray}
\theta \psi_1  &=& q  \psi_1 \theta \nonumber \\
\theta \psi_2  &=& q^2  \psi_2 \theta \\
\psi_2 \psi_1 &=& q \psi_1 \psi_2. \nonumber
\end{eqnarray}

\noindent
It becomes now straightforward to obtain the transformations on the
fsuperfield induced by fsusy transformations $\Phi(t,\theta) \longrightarrow
\Phi(t',\theta')$. Inserting the values of $t', \theta'$ obtained previously,
we
get the transformed fields :

\begin{eqnarray}
\Phi(t',\theta')&=& x(t') + q^2   \theta' \psi_2(t') +q^2 \theta'^2
 \psi_1(t') \nonumber \\
&=& x'(t) + q^2 \theta   \psi_2'(t)   +q^2 \theta^2   \psi_1'(t)
\nonumber \\
&=& x(t) + q^2 \epsilon \psi_2(t) +q^2 \epsilon^2 \psi_1(t) \\
&+& q^2 \theta ( \psi_2(t) + \epsilon^2 \dot x(t) -q \epsilon \psi_1(t))
\nonumber \\
&+&q^2 \theta^2( \psi_1(t)+ \epsilon \dot x(t) -q \epsilon^2  \dot \psi_2(t)),
\nonumber
\end{eqnarray}

\noindent
implying the fsuperfield components transformations\cite{fsusy1,fsusy2,fsusy3,
fsusygr,fsusy4}

\begin{eqnarray}
\delta_{\epsilon} x &=& q^2 \epsilon \psi_2 \nonumber \\
\delta_{\epsilon} \psi_2&=& -q \epsilon  \psi_1 \\
\delta_{\epsilon} \psi_1 &=& \epsilon \dot x. \nonumber
\end{eqnarray}

\noindent
It has to be underlined that the transformed fields
 $x'(t), \psi'_1(t), \psi'_2(t)$
do not satisfy the
same q-mutation relations as the initial ones. To cure this problem
in\cite{fsusy3}, a
cocycle was introduced to correct the statistics. However, there is no need
of such an object because the {\it only} fields that have  to fulfill the same
 q-mutations
as the initial ones are $x(t'), \psi_1(t'), \psi_2(t')$ and they {\it do}.
This is a quite general feature of quantum field theory. The reason why the new
fields $x'(t), \psi'_1(t), \psi'_2(t)$  do  not actually fulfill
the right q-mutation
relations is that we have broken down explicitly,
using Taylor expansion, the symmetry in the fsuperspace.

\noindent
The next step is to construct a representation  of the fsusy algebra acting on
$\Phi$, as well as a covariant derivative to establish    the action.
We  first need to recall some basic features of the derivation acting
on generalized Grassmann variables. This structure,
the q-deformed Heisenberg algebra, has been analyzed  in\cite{gca3}
 as well as its matrix representation\cite{gca3,hq}. It  admits in general
  $(n-1)$ derivatives, and we
note $\partial_\theta$ and $\delta_\theta$ the two derivatives of
the $n=3$ case which  satisfy

\begin{eqnarray}
\partial_\theta \theta - q \theta \partial_\theta &=& 1 \nonumber \\
\delta_\theta \theta -  q^2 \theta \delta_\theta  &=& 1 \nonumber \\
\partial_\theta^3=0 \  \ \ \  \delta_\theta^3&=&0 \\
\partial_\theta \delta_\theta = q^2 \delta_\theta \partial_\theta. && \nonumber
\end{eqnarray}

\noindent
Then let us introduce the two basic objets of the fsusy  $Q$ and $D$
the generator of fsusy and the covariant derivative respectively\cite{fsusy1,
fsusy2,fsusy3,fsusygr,fsusy4}

\begin{eqnarray}
Q &=& \partial_\theta + q \theta^2 \partial_t \nonumber \\
D &=& \delta_\theta + q^2 \theta^2 \partial_t.
\end{eqnarray}

\noindent
It can be checked explicitly that $D^3=Q^3=-\partial_t$ and
$QD=q^2 DQ$. A direct calculation proves that

\begin{eqnarray}
\Phi(t', \theta')&=& exp_{q^2}(\epsilon Q) \Phi(t,\theta) \\
\delta_\epsilon \Phi & =& \epsilon Q \Phi(t,\theta). \nonumber
\end{eqnarray}

\noindent
Using the fact that $D$ q-mute with $Q$ we have $\delta_\epsilon D \Phi =
D \delta_\epsilon \Phi$. Finally arguing that the $\theta^2$ component     of
$\Phi$ transforms like a total derivative we can take the opportunity to
construct
the action   by
taking the $\theta^2$ part of the action built in the fsuperspace.
 In  other words,
using the results on integration upon generalized Grassmann
variables\cite{intgca}
$\int d\theta = {d^n \over d\theta^n}
$ we have

\begin{eqnarray}
S &=& -{q^2 \over 2}\int  dt d \theta \dot \Phi D \Phi \nonumber \\
  &=& \int dt (\dot {x^2  \over 2} + {q^2 \over 2} \dot \psi_1 \psi_2
  -{q \over 2} \dot \psi_2 \psi_1).
 \end{eqnarray}

\noindent
So, from a  pure group theoretical approach one gets the basic action usually
used within the framework of fsusy\cite{fsusy1,fsusy2,fsusy3,fsusygr,fsusy4}.
 It can be pointed out that  this action
is real as it should be (for the q-mutators of the fields with the derivatives
see the appendix).

To gauge these symmetries { \it i.e}  to impose the invariance of the
action under local
diffeomorphism $t \longrightarrow t-f(t)$  and local fractional
supersymmetry { \it i.e} fractional supergravity (fsugra) we need to
introduce two {\it real} gauge   fields $e$ the einbein and  $\chi$ the
fractional gravitino (fgravitino), that couple    with their
associated conserved charged $H={1 \over 2} \dot x^2$ for the diffeomorphism
and $Q={q^2 \over 2} (\dot x \psi_2 +  {1 \over 2} \psi_1^2)$ for the
fsusy
(S. Durand in\cite{fsusy1}) respectively. Following the standard technics of
 gauge theory, noting
$\pi=\dot x$, $\pi_1={q^2 \over 2} \psi_2$ and $\pi_1={-q \over 2} \psi_1$
the conjugate momentum of $x,\psi_1, \psi_2$ and $H={1 \over 2} \pi^2$
the Hamiltonian,
 we have to
replace $L=\pi \dot x   + \dot \psi_1 \pi_1
+ \dot \psi_2 \pi_2 - H$  by

\begin{eqnarray}
L&=& \pi \dot x -{1 \over 2} e \pi^2 + \dot \psi_1 \pi_1
+ \dot \psi_2 \pi_2  \nonumber \\
&+& {q^2 \over 4} \chi (\pi \psi_2 +  {1 \over 2} \psi_1^2) \\
&+& {q \over 4} (\psi_2 \pi  +  {1 \over 2} \psi_1^2) \chi. \nonumber
\end{eqnarray}

\noindent
It is necessary to write, in the modified action, terms like
$q^2 \chi Q +  q Q^+ \chi$
to ensure the reallity of the new action. Taking the variation of the action
with respect to $\pi$, we get

\begin{equation}
\pi= { \dot x \over e} + {q^2 \over 2} { \chi \over e} \psi_2.
\end{equation}

\noindent
Inserting this value in the Lagrangian we finally
obtain

\begin{eqnarray}
L &=& {\dot x^2 \over 2e} + {q^2 \over 2} \dot \psi_1 \psi_2 -
{q \over 2} \dot \psi_2 \psi_1 \nonumber \\
&+& {q^2 \over 2}  \chi   ({ \dot x \over e} \psi_2 +
 {1 \over 2} \psi_1^2)
-{q^2 \over 4} { \chi^2  \over e} \psi_2^2.
\end{eqnarray}

\noindent
It is possible to rewrite this Lagrangian, introducing appropriate covariant
derivatives, similarly to  the spinning particles case\cite{sugra}

\begin{eqnarray}
D_t x &=& \dot x + {1 \over 2} q^2 \chi \psi_2 \nonumber \\
D_t \psi_1 &=& \dot \psi_1 + {3 \over 2} {1 \over e} q \chi^2 \psi_2   \\
D_t \psi_2 &=& \dot \psi_2 - {1 \over 2} q \chi \psi_1, \nonumber
\end{eqnarray}

\begin{eqnarray}
L &=& {1 \over 2e} D_t x D_t^+ x
+ {1 \over 4} \{(q^2 D_t \psi_1 \psi_2 + q \psi_2 D_t^+ \psi_1) \nonumber  \\
&-& {1 \over 4} (q D_t \psi_2 \psi_1 + q^2 \psi_1 D_t^+ \psi_2)   \\
&=& \int d \theta {-q^2 \over 4}
\{(\delta_\theta +q^2\theta^2 {1 \over e} D_t) \Phi D_t \Phi +h.c.\}
\nonumber
\end{eqnarray}

\noindent
This action is a reminiscence of an action built in a curved fractional
superspace in a way analogous to eq.(16). Like in the spinning particle
case\cite{wl1}, we
 introduce the  fractional
einbein $E_M^A$, and its inverse $E_A^M$,
    where
$M=\tau, \Theta, A = t,\theta$ are the  curved/tangent indices of the
fractional
superline, which control the invariance by translation ($X^M=(\tau,\Theta)
\longrightarrow X'^M=X^M-\xi^M(X) $).   We restrict oursleves to  affine
  transformations for $\Theta$, to ensure that it is
invertible. Using the definition for integration on generalized grassmann
variables we get $\int d\theta = \int d\theta' J^{-2}$ with $J$ the Jacobian
of the transformation. Then, following the way which leads to the
superdeterminant\cite{ber} the transformations

$$ \tau'=A\tau + B \Theta,\ \  \Theta'=C\tau + D\Theta,$$

\noindent
give the following fractional superdeterminant $det (A - BD^{-1}C) det^{-2} D$
(for arbitrary  $n$ we would have obtained $-n+1$ instead of $-2$). With such
a transformation we can build, using the fractional einbein,   an invariant
volume $S_qdet(E) = (E^t_\tau - E^t_\Theta E^\Theta_\theta E^\theta_\tau)
(E^\Theta_\theta)^2$. Doing so, we obtain the action

\begin{equation}
S=-{ q^2 \over 4} \int dt d\theta S_qdet(E)( E^M_t \partial_M \Phi E^N_\theta
\partial_N \Phi + h.c),
\end{equation}

\noindent
with a huge invariance corresponding to the reparametrization of
the fractional superline and possibly to transformations on the metric
like in the spinning case\cite{wl1}.
The fsugra transformations,
would correspond to a subset of these transformations with a special
constrained
choice of the parameters. Due to the Nother procedure
the action (19) shall be invariant under this  subset of transformations.
Of course, one could whish to have an explicit formulation of the fsugra
transformations. But this will be devoted to a future publication. To obtain
the analogue of the Dirac equation, there is no need to know these
transformation laws. What we need is just the local action (19).
  This action
is invariant under the local diffeomorphism

\begin{eqnarray}
\delta_fx &=&f \dot x \nonumber \\
\delta_f \psi_1 &=&f \dot \psi_1 \nonumber \\
\delta_f \psi_2 &=& f \dot \psi_2   \\
\delta_fe &=& f \dot e + \dot f e \nonumber \\
\delta_f \chi &=&f \dot \chi + \dot f \chi, \nonumber
\end{eqnarray}

\noindent
and under fsugra transformations.\\

\noindent
{\bf  II.Dirac quantization}\\

Having obtained the full action which is
 invariant under fsupergravity transformation
and one dimensional diffeomorphism, we are now in a position to quantize our
theory. As for the spinning particles\cite{wl1}, we are typically with a system
which
presents constraints. We have two second class constraints because the momenta
of $\psi_1, \psi_2$ are not independent of the fields themselves

\vfill \eject
\begin{eqnarray}
\Xi_1= \pi_1-{q^2 \over 2} \psi_2 &=& 0 \\
\Xi_2= \pi_2+{q \over 2} \psi_1 &=& 0,  \nonumber
\end{eqnarray}

\noindent
and two first class constraints resulting of the gauge invariance of
the $1D$ fsugra and  diffeomorphism

\begin{eqnarray}
{\delta S \over \delta e} &=& H = {1 \over 2} \pi^2 = 0 \\
{\delta S \over \delta \chi} &=& Q = {q^2 \over 2}( \pi \psi_2 + {1 \over 2}
\psi_1^2)=0, \nonumber
\end{eqnarray}

\noindent
so the einbein and the fgravitino just appear as Lagrange multiplier for
the constraints. The quantization of a theory with constraints has been
studied by Dirac\cite{Dirac}, and a different treatment as to be implemented
for
first and second class constraints. For the first class ones we have to
substitute the Poisson bracket to the Dirac one, and so the first step is to
define an appropriate Poisson bracket for variables that q-mute. This
can be done by  using the q-symplectic metric\cite{q-part} $\Omega$ or by
noting that $\psi_1, \psi_2$ are in a representation of the quantum
 hyperplane\cite{q-plane}
which admits a $R-$matrix convenient for such a construction
( see for example\cite{q-group}). Recall that\cite{q-part}

\begin{equation}
\Omega=\pmatrix{0&1&0&0& \cr
                -q&0&0&0&\cr
                0&0&0&1&\cr
                0&0&-q^2&0&},
\end{equation}
\noindent
where the indices $1,2,3,4$ are respectively for $\psi_1,\pi_1,\psi_2,\pi_2$
and that the $R-matrix$ stating $\psi_2 \psi_1 = q  \psi_1 \psi_2$
is\cite{q-plane}

\begin{equation}
R=\pmatrix{q^2&0&0&0& \cr
                0&q^2 -q&1&0&\cr
                0&1&0&1&\cr
                0&0&0&q^2&}.
\end{equation}

\noindent
With the use of  $\Omega$  \cite{q-part} or  $R$ \cite{q-group} we obtain

\begin{eqnarray}
\{A,B\}_{P.B.} &=& {\partial A \over \partial x} {\partial B \over \partial
\pi}
              -{\partial A \over \partial \pi} {\partial B \over \partial x}
              \nonumber \\
 &+& {\partial A \over \partial \psi_1} {\partial B \over \partial \pi_1}
 - q {\partial A \over \partial \pi_1} {\partial B \over \partial \psi_1} \\
 &+&  {\partial A \over \partial \psi_2} {\partial B \over \partial \pi_2}
 - q^{-1} {\partial A \over \partial \pi_2} {\partial B \over \partial \psi_2}.
 \nonumber
 \end{eqnarray}

\noindent
With this definition we can check explicitly that $\Xi_1,\Xi_2$ are
second class constraints by calculating  the
algebra of the constraints

\begin{equation}
\{\Xi_i,\Xi_j\}_{P.B.} = C_{ij} =
                        \pmatrix{0&-q^2 & \cr
                                 q&0&}.
\end{equation}
\noindent
So following Dirac we define the Dirac bracket

\begin{equation}
\{A,B\}_{D.B} = \{A,B\}_{P.B} - \{A,\Xi_i\}_{P.B} C^{-1}_{ij}
\{\Xi_j,B\}_{P.B}.
\end{equation}

\noindent
When we calculate the Dirac bracket of two $\psi$'s and substitute the
Dirac bracket by the q-mutator we obtain the quantized variables which
satisfy

\begin{eqnarray}
\psi_1 \psi_1 - \psi_1 \psi_1&=&0 \nonumber \\
\psi_2 \psi_2 - \psi_2 \psi_2&=&0 \\
\psi_2 \psi_1 - q\psi_1 \psi_2&=& {- q^2}. \nonumber
\end{eqnarray}

\noindent
This result is in exact accordance with the fact that the conjugate momentum
of $\psi_1$ is $ {q^2 \over 2} \psi_2$ and it is wellknown that they have
to belong to the q-deformed Heisenberg algebra\cite{gca3} ( see for instance
 the
q-mutator of $\theta, \partial_\theta$).
With the quantized variables we directly check that the algebra (of the
first class constraints) closes and we have $Q^3 = H$ so the first class
constraints are imposed upon the physical states

\begin{equation}
H | \lambda_{phys}> = Q |\lambda_{phys}>  = 0.
\end{equation}

\noindent
To interpret these two equations we have   to build the corresponding
Fock space, but first
we would like to have a formalism adapted to space-time {\it i.e.} when
the variables carry space-time indices.  So the space-time is just the
target space in which the world-line is embedded. The full action (19),
besides   its $1D$ dimensional invariance, is  imposed to be  $D-$dimensional
Poincar\'e invariant. Noting $x^\mu, \psi_1^\mu,\psi_2^\mu$ the basic fields
which are in the vectorial representation of the Poincar\'e group
and $\eta_{\mu \nu}$ the Minkowski metric $\eta = diag(1,-1,\cdots,-1)$
we have

\begin{eqnarray}
L &=& {\dot x^\mu \dot x^\nu  \over 2e} \eta_{\mu \nu}
 + {q^2 \over 2} \dot \psi_1^\mu \psi_2^\nu \eta_{\mu \nu} -
{q \over 2} \dot \psi_2^\mu \psi_1^\nu \eta_{\mu \nu} \nonumber \\
&+& {q^2 \over 2} \chi ({ \dot x^\mu \over e}   \psi_2^\nu \eta_{\mu \nu}+
{1 \over 2} \psi_1^\mu \psi_1^\nu \eta_{\mu \nu} )
-{q^2 \over 4} { \chi^2  \over e} \psi_2^\mu \psi_2^\nu \eta_{\mu \nu}.
\end{eqnarray}

\noindent
Everything we have done up to now is suitable except that the product is
replaced by  a scalar product. The first question which arises concerns the
q-mutation relations between the different components. Remember that we
have various
constraints

-(i) the fields have to be in a vectorial representation of the Poincar\'e
group

-(ii) the fractional supercharge has to close the algebra:
$$Q^3=(\pi^\mu \psi_2^\nu \eta_{\mu \nu} +
{1 \over 2} \psi_1^\mu \psi_1^\nu \eta_{\mu \nu})^3   =
  {1 \over 2} \pi^\mu \pi^\nu \eta_{\mu \nu}, $$

\noindent
with $\pi_\mu$ the conjugate momentum of $x^\mu$.
This last relation is very strong and following the results of\cite{gca2} on
the linearization of polynomial( especially theorem 1.2 and its corollary
and proposition 2.1, 2.2) the various components $\psi_1^\mu, \psi_2^\nu$
have to q-mute (see appendix for the q-mutations). However,  those
q-mutation relations are not stable under $SO(1,D-1)$ but only through
the quantum group\cite{q-plane} $GL_q(D)$. So at  a first glance it seems that
(i) and (ii) are incompatible. However there is no need to impose the stability
of the q-mutation relations under Lorentz transformations; so  we just set the
q-mutator

\begin{eqnarray}
\psi_a^\mu  \psi_b^\nu = q \psi_b^\nu \psi_a^\mu, \ \ (a,\mu)< (b,\nu)
\cdots a,b=1,2; \mu,\nu=0,1, \cdots D-1,
\end{eqnarray}

\noindent
(the two possibilities of lexicographical order are defined in the appendix) in
one special frame, called the q-frame or the q-gauge and by covariance if in
this frame we have $Q^3=H$ it will be the case in {\it any} frame.
It can be pointed out that a similar property appears in Yang-Mills theory,
where the commutation relations of the components of the gauge field are
not preserved under Lorentz transformations. Moreover, we see that the
variables
then obtained are not in a representation of the quantum
hyperplane\cite{q-plane}
because
we have $\psi_2^\mu \psi_1^\mu  - q \psi_1^\mu \psi_2^\mu = -q^2$.

With the $2D$ variable $\psi_a^\mu$ we can, in this q-frame, define $D$
series of q-deformed oscillators\cite{q-osc.} $a_\mu, a^+_\mu$

\begin{eqnarray}
&&a_\mu^3 = a_\mu^{+3} = 0 \nonumber \\
&&a_\mu a^+_\mu -q^{-1} a^+_\mu a_\mu = q^{N_\mu} \nonumber \\
&&a_\mu a_\nu = q a_\nu a_\mu,\ \ \mu<\nu   \\
&&a^+_\mu a^+_\nu = q a^+_\nu a^+_\mu,\ \ \mu<\nu \nonumber \\
&&a^+_\mu a_\nu = q^{+/-} a_\nu a^+_\mu. \nonumber
\end{eqnarray}

\noindent
Where the ${+/-}$, in the last equation, corresponds to the two possible
orderings (see appendix) and $N_\mu$ is the $\mu-$th number operator. With
the $a^+_\mu $ we can build a  $3D$ Fock space. If we note $|0>$ the
vacuum annihilated by all the  $a$'s we have

\begin{equation}
|\lambda_{phys} > = (a_0^+)^{\alpha_0} \cdots (a_{D-1}^+)^{\alpha_{D-1}} |0>,
\ \ \ \alpha_{D-1}, \cdots  \alpha_0=0,1,2,
\end{equation}

\noindent
and the relativistic wave $<x|\lambda_{phys}> =\lambda(x)$ satisfies the
wave equations

\begin{eqnarray}
\partial_\mu \partial^\mu \lambda(x) &=& 0 \nonumber \\
( i \psi_2^\mu \partial_\mu+ {1 \over 2}  \psi_1^\mu\psi_{1\mu} )\lambda(x)
&=& 0.
\end{eqnarray}

\noindent
Using a result of\cite{gca2}, that is, given a set of $k$ operators
$A_1, \cdots A_k$
satisfying $A_i A_j = q A_j A_1, i < j$ then $(\sum\limits_{i=1}^k A_i)^3 =
(\sum\limits_{i=1}^k (A_i)^3)$,
 we can easily prove that
$$(i \psi_2^\mu \partial_\mu+ {1 \over 2}  \psi_1^\mu\psi_{1\mu})^3=
-{1 \over 2} \partial_\mu \partial^\mu.$$
This new operator can be seen as a cubic root of the d'Alembertian operator
extending the Dirac equation to the fractional Dirac equation (fDirac),
although the first equation tells us that we have a massless particle.
Looking to equations (35), we just see that these new states are not in a
representation of the permutation group but of the braid group. So
we obtain states which constitute neither fermions or bosons nor parafermions
or parabosons\cite{para} but describe alternative statistics. Do we get anyons,
but without Chern-Simon\cite{anyons} terms , or fractional statistics?
This is an open
question. Of course such a representation is allowed for $D \leq 3$,
to prove that
$D$ is constrainted is still an open question. Some hint,
to understand the meaning of (37),
can be given. We can include, in our global Lagrangian,
an additional term representing \\

  - the interaction with an electromagnetic field. Note $g$ the coupling
  constant,
$A_\mu(\Phi)$  the gauge field defined in the fsuperspace and $F_{\mu \nu}(x)$
the electromagnetic tensor

\begin{eqnarray}
  L_{e.m}& = -{q^2 \over 2} & \int d \theta (g A^\mu(\Phi) D \Phi_\mu  + h.c.)
   \nonumber \\
   &=& g \dot x_\mu A^\mu+ {g \over 2}  (q \psi_2^\mu \psi_1^\nu
   +q^2 \psi_1^\mu \psi_2^\nu) F_{\nu \mu} \\
   &+& {g \over 4}(q^2 \psi_2^\alpha \psi_2^\beta \psi_2^\mu +
                   q   \psi_2^\mu \psi_2^\beta \psi_2^\alpha)
                   \partial_\alpha  \partial_\beta A_\mu ;
                     \nonumber
\end{eqnarray}

  - as well as a coupling to the gravitational field. Note $g_{\mu \nu}(\Phi)=
     \eta_{\mu \nu}  + h_{\mu \nu}(\Phi)$ the metric defined  in the
     fractional superspace

\begin{eqnarray}
L_{grav.} &=& {-q^2 \over 4} \int d \theta (\dot \Phi^\mu D \Phi^\nu +h.c.)
  h_{\mu \nu}(\Phi) \nonumber \\
 &=& {1 \over 2} ( \dot x^\mu \dot x^\nu +q^2 \dot \psi_1^\mu \psi_2^\nu
 -q \dot \psi_2^\mu \psi_1^\nu)  h_{\mu \nu} \\
 &+&{1 \over 4}  ((-q^2 \dot x^\mu \psi_1^\nu \psi_2^\alpha
                   + q \dot x^\mu \psi_2^\nu \psi_1^\alpha
                   + q \dot \psi_2^\mu \psi_2^\nu \psi_2^\alpha)
                   \partial_\alpha h_{\mu \nu} +h.c.)   \nonumber \\
 &+&{1 \over 8} (q \dot x^\mu \psi_2^\nu \psi_2^\alpha  \psi_2^\beta +h.c.)
                   \partial_\alpha \partial_\beta h_{\mu \nu} +h.c.).
 \nonumber
\end{eqnarray}

\noindent

 As a final remark we should notice
that all this has nothing to do with the q-deformed spinning
particles\cite{q-part}.

\vskip .5 truecm

\noindent
{\bf  III. Concluding  remarks}\\

In a way similar to the one which has led to the description of the
spinning particles, we have obtained, using local fsusy, a relativistic wave
equation which describes states endowed with alternative statistics. The
interpretation of what kind of particle we are describing is still an open
question, just as the fact that the dimension $D$ cannot excess three.
The case of the  massive particles can be undertaken in a way similar to
the one  that has been
introduced in the framework of $1D$ susy\cite{wl1}, {\it i.e.} by the
introduction of an auxiliary field. It should be interesting to extend
all this formalism for any $n$. However, for that purpose we need a systematic
way to obtain the local action, introducing a curved fractional superspace
( see eq. (22)) and paying attention to the symmetries of these action
like for instance
   in the spinning particle case\cite{wl1}.

We have obtained these results using a one dimensional lagrangian
formalism. However it should be possible to build similar actions for $D=2,3$.
The case $D=2$ has been considered in an heterotic way\cite{fsusy3}, and
in connection with string\cite{naka}. Those two actions should be analyzed
within the framework of the
 fractional Virasoro\cite{fsusy4} algebra and   in connection
with representation of the two dimensional Virasoro algebra.

As a final remark we just want to say that it is interesting to study
all these kind of algebras in order to see how they can have connections with
space-time properties.\\

\vfill \eject
\noindent
{\bf Acknowledgement}
\vskip .5 truecm
We would like to acknowledge J. -L. ~Jacquot and J.~ Polonyi for useful
remarks and encouragments.

\vskip .5 truecm
 \noindent
{\bf  Appendix}\\
\vskip .5 truecm

\noindent

In this appendix we just want to set up the q-mutation relations between
 the various
fields and  to see how they are arbitrary.
We have two types of q-mutation relations, when the space-time indices are not
involved and among the various components. First write the q-mutation
without the space-time index.

$$\theta x = x\theta  \eqno(A.1)$$
$$\theta \psi_1 = q \psi_1 \theta  \eqno(A.2)$$
$$\theta \psi_2 = q^2 \psi_2 \theta,  \eqno(A.3)$$

$$\partial_\theta x = x \partial_\theta \eqno(A.4)$$
$$\partial_\theta \psi_1 = q^2 \psi_1 \partial_\theta \eqno(A.5)$$
$$\partial_\theta \psi_2 = q \psi_2 \partial_\theta. \eqno(A.6)$$

\noindent
We have the same q-mutation with $\partial_\theta$ replaced by
$\delta_\theta$.

$$\epsilon x = x \epsilon \eqno(A.7)$$
$$\epsilon\psi_1 = q \psi_1 \epsilon \eqno(A.8)$$
$$\epsilon\psi_2 = q^2 \psi_2 \epsilon, \eqno(A.9)$$

$$\chi x = x\chi \eqno(A.10)$$
$$\chi\psi_1 = q \psi_1 \chi \eqno(A.11)$$
$$\chi\psi_2 = q^2 \psi_2 \chi. \eqno(A.12)$$

\noindent
The relations (A.1-3) come from the definition of the
grading, ($x,\psi_1,\psi_2$)
are of grading $0,1,2$ respectively.
 Of course the derivatives
have  the  q-mutators with  $q \longrightarrow q^{-1}$.
 We could have chosen equivalently the other
cubic root $q^2$, which would have led to a substitution of $q \longrightarrow
q^{-1}$ in {\it all the q-mutation relations}.

\noindent
(A.7-9) are consequences of (A.1-3) due to the translation $\theta'=\theta+
\epsilon$ in the fsuperspace, they are also compatible with the fact that
the variations on the fields under fsusy are real.
\noindent
(A.10-12) are consequences  of the fsugra transformation ( from
Nother theorem it is known that $\delta_\epsilon \chi \sim \dot \epsilon)$.

$$\theta \epsilon  = q \epsilon \theta \eqno(A.13)$$
$$\theta \chi  = q\chi \theta  \eqno(A.14)$$
$$\epsilon \chi = q \chi \epsilon, \eqno(A.15)$$

\noindent
(A.13) is imposed in order that the fsusy variations in the fsuperspace are
real, (A.14-15) results from (A.13).

\noindent
$$\partial_\theta \theta -q \theta \partial_\theta =1 \eqno(A.16)$$
$$\delta_\theta \theta -q^2 \theta \delta_\theta =1 \eqno(A.17)$$
$$\partial_\theta \delta_\theta = q^2 \delta_\theta \partial_\theta,
 \eqno(A.18)$$

\noindent
by definition of the derivative in the q-deformed Heisenberg algebra.

$$\psi_2 \psi_1 = q \psi_2 \psi_1 \eqno(A.19)$$
$$ \dot \psi_1 \psi_1 = q^{a_1} \psi_1 \dot \psi_1 \eqno(A.20)$$
$$ \dot \psi_2 \psi_2 = q^{a_2} \psi_2 \dot \psi_2, \eqno(A.21)$$

\noindent
(A.19) ensures the reality of the Lagrangian.
  {\it
A priori} from $\psi_i^3=0$ we deduce that
$ \dot \psi_i \psi_i = q^{a_i} \psi_i \dot \psi_i$ but the choice of $a_i$'s
are arbitrary.
Finally note that the derivative of the fields satisfy the same q-mutation
relations as the field themselves.\\
\noindent
After quantization  (A.19) becomes

$$ \psi_2 \psi_1 - q \psi_1  \psi_2 = -q^2. \eqno(A.22)$$

To conclude we  write the q-mutation relations when the space-time
indices are involved

$$\psi_a^\mu \psi_b^\nu = q \psi_b^\nu \psi_a^\mu, \ \ \ (a,\mu) < (b,\nu),
\eqno(A.23)$$

\noindent with $a,b=1,2$ and $\mu,\nu=0,1,2$. Two possible orderings are
allowed

(1) $(a,\mu) < (b,\nu)$ if $\mu< \nu$ else  $\mu= \nu $ and $a< b$;

(2) $(a,\mu) < (b,\nu)$ if $a< b$ else  $a= b $ and $\mu< \nu$.

\noindent
And only (2) allows  a matrix representation. With the notations of\cite{hq}
we have ($D=3$ and for an Euclidian space, or after a Wick rotation)

$$\psi_1^0=\theta \otimes \sigma_3 \otimes \sigma_3 \eqno(A.24)$$
$$\psi_1^1=I_3 \otimes \theta  \otimes \sigma_3 \eqno(A.25)$$
$$\psi_1^2=I_3 \otimes I_3 \otimes \theta,   \eqno(A.26)$$

$$\psi_2^0= -q^2 \partial_\theta \otimes \sigma_3^2 \otimes \sigma_3^2
\eqno(A.27)$$
$$\psi_2^1=-q^2. I_3 \otimes \partial_\theta  \otimes \sigma_3^2 \eqno(A.28)$$
$$\psi_2^2=-q^2. I_3 \otimes I_3 \otimes \partial_\theta,    \eqno(A.29)$$

\noindent
with

$$\theta = \pmatrix{0&0&0& \cr
                    1&0&0& \cr
                    0&1&0& \cr}$$

$$\partial_\theta = \pmatrix{0&1&0& \cr
                             0&0&1+q& \cr
                             0&0&0& \cr}$$

$$\sigma_3 = \pmatrix{1&0&0& \cr
                      0&q&0& \cr
                      0&0&q^2& \cr}.$$

\noindent
$\theta,\partial_\theta$ are the matrices which appear in the literature
of the $q-$deformed Heisenberg algebra\cite{gca3,hq}, $\sigma_3$ is one of the
basic matrix
appearing wi\-thin the framework of generali\-zed Clifford algebra(
 see\cite{gca2}
and
references therein) and $I_3$ is the $3 \times 3$ indentity matrix.


\end{document}